\documentclass[preprint,5p,times]{elsarticle}

\usepackage{graphicx}
\usepackage{dcolumn}
\usepackage{amsmath} 
\usepackage{physics}
\usepackage{subfloat}
\usepackage{bm}
\usepackage{xcolor} 
\biboptions{numbers,sort&compress}

\newcommand{\nuc}[2]{$^{#1}$#2}

\newcommand{\zps}{$0^{+}$ states}
\newcommand{\dpsr}{d($^{95}$Sr,p)}

\journal{Physics Letters B}

\begin{document}
\begin{frontmatter}

\title{Shape Coexistence and Mixing of Low-Lying $0^+$ States in $^{96}$Sr}

\author[ubc,tr]{S.~Cruz}
\author[tr]{P.C.~Bender}
\author[ubc,tr]{R.~Kr\"{u}cken}
\author[ut,cmu]{K.~Wimmer\corref{cor1}}
\author[tr]{F.~Ames}
\author[sfu]{C.~Andreoiu}
\author[sm]{R.A.E.~Austin}
\author[cmu]{C.S.~Bancroft}
\author[csm]{R.~Braid}
\author[tr]{T.~Bruhn}
\author[su]{W.N.~Catford}
\author[tr]{A.~Cheeseman}
\author[sfu]{A.~Chester}
\author[sfu]{D.S.~Cross}
\author[yo]{C.Aa.~Diget}
\author[to]{T.~Drake}
\author[tr]{A.B.~Garnsworthy}
\author[tr]{G.~Hackman}
\author[sm,tr]{R.~Kanungo}
\author[su]{A.~Knapton}
\author[ir,tr]{W.~Korten}
\author[csm]{K.~Kuhn}
\author[tr]{J.~Lassen}
\author[tr]{R.~Laxdal}
\author[tr]{M.~Marchetto}
\author[su,lpc]{A.~Matta}
\author[tr]{D.~Miller}
\author[tr]{M.~Moukaddam}
\author[lpc]{N.A.~Orr}
\author[cmu]{N.~Sachmpazidi}
\author[sm,tr]{A.~Sanetullaev}
\author[gu]{C.E.~Svensson}
\author[cmu]{N.~Terpstra}
\author[tr]{C.~Unsworth}
\author[sfu]{P.J.~Voss}

\address[ubc]{Department of Physics and Astronomy, University of British Columbia, Vancouver, BC V6T 1Z4, Canada}
\address[tr]{TRIUMF, Vancouver, BC V6T 2A3, Canada}
\address[ut]{Department of Physics, The University of Tokyo, Hongo, Bunkyo-ku, Tokyo 113-0033, Japan}
\address[cmu]{Department of Physics, Central Michigan University, Mt Pleasant, MI 48859, USA}
\address[sfu]{Department of Chemistry, Simon Fraser University, Burnaby, BC V5A 1S6, Canada}
\address[sm]{Department of Astronomy and Physics, Saint Mary's University, Halifax, NS B3H 3C2, Canada}
\address[csm]{Department of Physics, Central Michigan University, Mt Pleasant, MI 48859, USA}
\address[su]{Department of Physics, University of Surrey, Guildford, Surrey, GU2 7XH, United Kingdom}
\address[yo]{Department of Physics, University of York, York, YO10 5DD, United Kingdom}
\address[to]{Department of Physics, University of Toronto, Toronto, ON M5S 1A7, Canada}
\address[ir]{IRFU, CEA, Universit\'{e} Paris-Saclay, F-91191 Gif-sur-Yvette, France}
\address[lpc]{LPC, ENSICAEN, CNRS/IN2P3, UNICAEN, Normandie Universit\'{e}, 14050 Caen cedex, France}
\address[gu]{Department of Physics, University of Guelph, Guelph, ON, N1G 2W1, Canada}
\date{\today}

\begin{abstract}
  The low energy excited $0_{2,3}^+$ states in $^{96}$Sr are amongst the most prominent examples of shape coexistence across the nuclear landscape. In this work, the neutron $[2s_{1/2}]^2$ content of the $0_{1,2,3}^+$ states in $^{96}$Sr was determined by means of the \dpsr\ transfer reaction at the TRIUMF-ISAC2 facility using the SHARC and TIGRESS arrays. Spectroscopic factors of 0.19(3) and 0.22(3) were extracted for the $^{96}$Sr ground and 1229~keV $0^+$ states, respectively, by fitting the experimental angular distributions to DWBA reaction model calculations. A detailed analysis of the $\gamma$-decay of the isomeric $0_3^+$ state was used to determine a spectroscopic factor of 0.33(13). The experimental results are compared to shell model calculations, which predict negligible spectroscopic strength for the excited $0^+$ states in \nuc{96}{Sr}. The strengths of the excited $0_{2,3}^+$ states were also analyzed within a two-level mixing model and are consistent with a mixing strength of $a^2$=0.40(14) and a difference in intrinsic deformations of $|\Delta \beta|=0.31(3)$. 
  These results suggest coexistence of three different configurations in \nuc{96}{Sr} and strong shape mixing of the two excited \zps.
\end{abstract}


\end{frontmatter}

Describing the shape evolution of atomic nuclei presents a challenge to modern nuclear structure theory. The shape of the nucleus is a result of a delicate interplay between macroscopic, liquid drop-like and microscopic shell structure effects. Nuclei with a closed shell configuration are spherical in their ground states, but away from magic numbers deformed ground states are observed.
The degree of deformation results from the interaction between protons and neutrons depend on the exact occupation of single-particle orbitals near the Fermi surface. 
Therefore, small changes in the nucleon number can lead to rapid changes in both the magnitude and type of deformation. One of the most dramatic examples is the region of neutron-rich Zr ($Z=40$) and Sr ($Z=38$) isotopes. 
While the properties of Zr and Sr nuclei with $N\leq 58$ indicate spherical ground states, with the addition of just two neutrons the ground states become strongly deformed for $N=60$ and beyond.
The nuclei at this shape transition around $N=60$ exhibit shape coexistence~\cite{Heyde2011} with low-lying excited deformed (spherical) states for nuclei with $N\leq 58$ ($N\geq 60$).
The sudden onset of deformation in \nuc{100}{Zr} has been explained by the strong residual interaction between the proton-neutron spin-orbit partner orbitals $\pi[ 0g_{9/2}]$ and $\nu [0g_{7/2}]$. While in an independent particle picture the $\pi[ 0g_{9/2}]$ orbital is completely empty in Zr, adding neutrons to the $\nu [0g_{7/2}]$ orbital enables the promotion of protons from the lower lying orbitals to the deformation-driving $\pi [0g_{9/2}]$ orbital~\cite{FP1977,FP1979}. The resulting sudden transition from spherical to deformed ground states and the emergence of shape-coexisting states in the vicinity of $N=60$ and $Z=40$ has been a subject of considerable interest for many years, both theoretically~\cite{Arseniev1969,FP1977,FP1979,Kumar1985,Michiaki1990,Skalski1993,Baran1995,Lalazissis1995,Skalski1997,Holt2000,Zhang2006,Rza209,RodGuz2010a,Sieja2009,RodGuz2010, Liu2011,Mei2012,Xiang2012,Petrovici2012,Togashi2016} and experimentally~\cite{Kratz1982,Buchinger1990,Mach1991,Wu2004,Park2016,Clement2016,Clement2016a,Regis2017}.

In Sr, the shape transition is evident through measurements of binding energies~\cite{Wang2016}, charge radii~\cite{Buchinger1990}, excitation energies and quadrupole transition probabilities of low-lying states~\cite{Mach1991,Clement2016,Clement2016a}. Low-lying \zps\ with strong electric monopole ($E0$) transitions between them indicate the coexistence of states with different intrinsic deformations or the occurrence of strongly mixed configurations~\cite{Wood1999}. In $^{96}$Sr ($N=58$) the low-lying 1229 and 1465~keV $0_{2,3}^+$ states are associated with shape-coexistence, as evidenced by the very strong monopole transition strength  $\rho^2(E0)=0.185(50)$~\cite{Jung80} between them. However, the measured lifetimes and extracted $E0$ and $E2$ transition strengths for the decay of the $0_2^+$ and $0_3^+$ states do not allow the conclusive determination of the mixing amplitudes between the two excited $0^+$ states or their relative deformation.
The spectroscopic quadrupole moments of the $2^+_1$ state in \nuc{96}{Sr} and the $2^+_2$ state in \nuc{98}{Sr} were found to be very similar~\cite{Clement2016,Clement2016a}, indicating that these states possess similar underlying structure. The mixing of the two coexisting shapes in $^{98}$Sr is weak despite their proximity~\cite{Park2016,Clement2016}.

Several theoretical studies have provided information on electromagnetic transition probabilities in $^{96}$Sr. The low-lying $0^+$ states in $^{96}$Sr have been studied using the complex excited VAMPIR method with a realistic effective interaction in a large model space~\cite{Petrovici2012}. In that work, the lowest three $0^+$ states were associated with triple shape coexistence of spherical, prolate, and oblate configurations. However, the strong $E0$ transition between the $0_2^+$ and $0_3^+$ states in $^{96}$Sr was not reproduced. The beyond mean field calculations with the Gogny D1S interaction of Refs.~\cite{Clement2016,Clement2016a} predict two excited bands in $^{96}$Sr with only moderate deformation but substantial triaxiality.
Recently, Monte-Carlo shell model calculations in a large model space were able to reproduce the energy levels and $B(E2)$ values of low-lying states in the Zr isotopic chain~\cite{Togashi2016}. When extended to the Sr isotopes~\cite{Regis2017}, these calculations describe well the level schemes and $B(E2)$ values of \nuc{94}{Sr} and \nuc{98}{Sr} but predict already substantial deformation for \nuc{96}{Sr}. For all these theoretical calculations the agreement with the experimental information is much better in \nuc{98}{Sr} and \nuc{98}{Zr} than \nuc{96}{Sr}.

In this Letter, we report on the first investigation of the low-lying $0^+_{1,2,3}$ states in $^{96}$Sr through the \dpsr\ transfer reaction at 5.5 $A$MeV in inverse kinematics, providing crucial insights into the shape mixing and differences in deformation of the coexisting shapes in $^{96}$Sr.
In contrast to the experimental work performed to date, we investigate the shape transition in \nuc{96}{Sr} from a different perspective and study the single-particle properties. Since the \nuc{95}{Sr} $1/2^+$ ground state is dominated by the $[2s_{1/2}]$ spherical single particle configuration, as we demonstrated in the present experimental campaign~\cite{Cruz2018}, the $\ell=0$ transfer in the \dpsr\ reaction probes the component of the $[2s1/2]^2$ configuration in the $0^+$ state wave functions.
Our work allowed for the first time to establish the mixing between the different shapes in \nuc{96}{Sr}. The results indicate a strong mixing of the two excited \zps\ while the weak population of the grounds state hints to the coexistence of three shapes in \nuc{96}{Sr}.

The present experiment was performed at the TRIUMF-ISAC2 facility~\cite{TRIUMF} where a $^{95}$Sr beam was produced by impinging a 480 MeV proton beam with an intensity of 10~$\mu A$ on a UC$_x$ target. The neutron-rich Sr isotopes were produced through uranium fission, were laser ionized, mass separated and transported to a charge state booster~\cite{CSB}. The beam (Q=16$^+$) was transported to the ISAC2 facility where its kinetic energy was increased to 5.5 $A$MeV using the superconducting linear accelerator~\cite{TRIUMF}. 
The post-accelerated $^{95}$Sr beam was delivered for approximately 2.5 days with an average intensity of $1.5\times10^6$ particles per second. The beam impinged upon a 0.44(4) mg/cm$^2$, 92(1)\% deuterated polyethylene (CD$_2$) target, mounted in the centre of the SHARC silicon detector array~\cite{SHARC}. SHARC (Silicon Highly-segmented Array for Reactions and Coulex) is a compact arrangement of double-sided silicon strip detectors which is optimized for high geometrical efficiency and excellent spatial resolution, with $\Delta \theta_\text{lab} \geq 1 ^\circ$. The SHARC array was surrounded by the TIGRESS $\gamma$-ray detector array, with 12 HPGe Compton-suppressed clover detectors arranged in a compact hemispherical arrangement with approximately 50\% of $4\pi$ geometrical coverage~\cite{TIGRESS}. The beam composition was measured periodically throughout the experiment using a Bragg ionization detector~\cite{TBRAGG}, which was positioned adjacent to the TIGRESS experimental station. The composition of the $A=95$ beam was 95(3)\% $^{95}$Sr.

\begin{figure}[h!]
\includegraphics[width=\columnwidth]{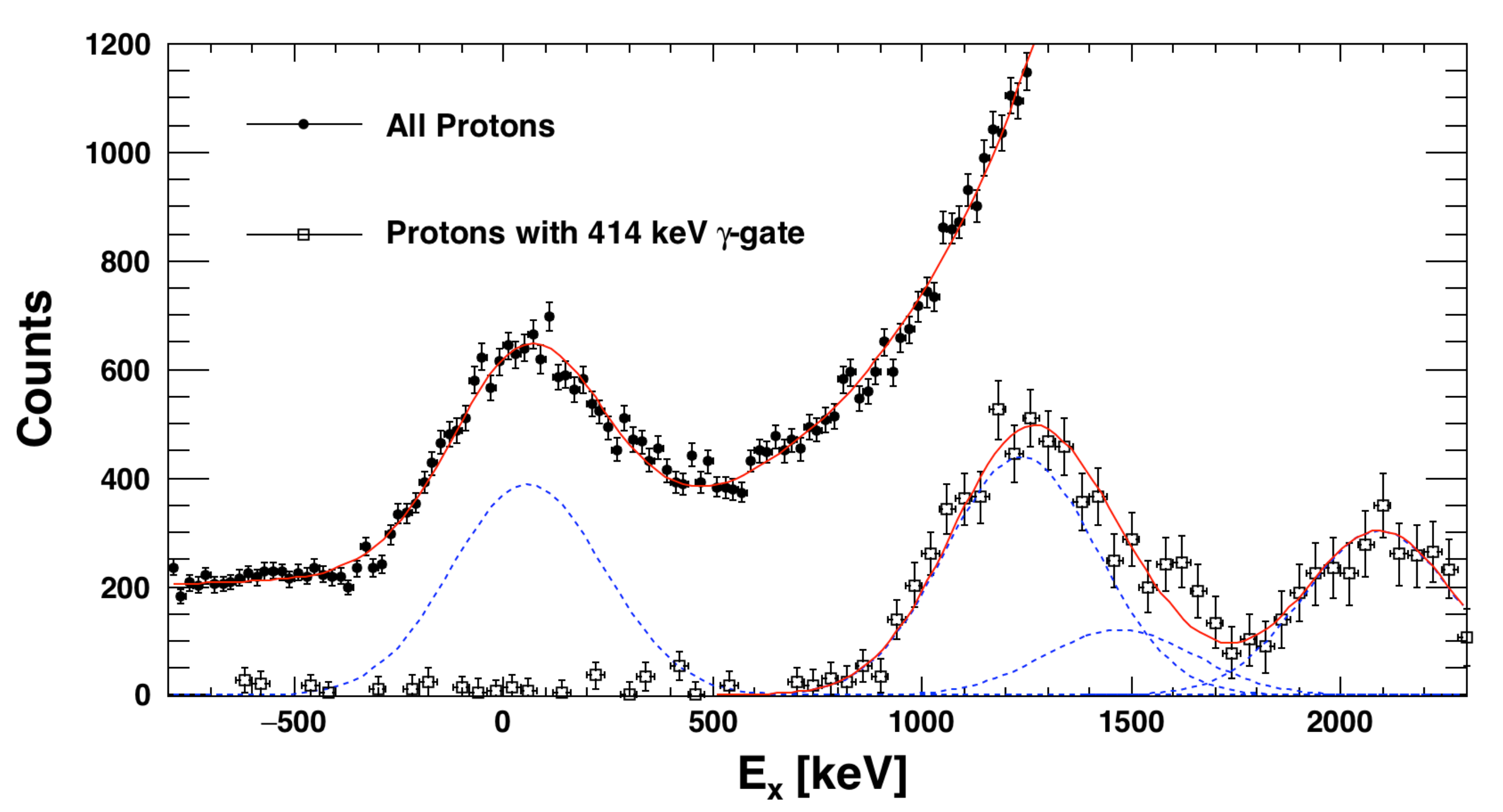}%
\caption{Excitation energy spectrum of \nuc{96}{Sr} obtained from energies and angles of protons emitted at backward laboratory angles.
The total spectrum (filled points) includes the information from all measured protons in the angular range $\theta_\text{lab}>90^\circ$ while the gated spectrum (open points), corrected for the efficiency of TIGRESS, is additionally gated on coincident 414~keV $\gamma$-rays from the $0_2^+ \rightarrow 2_1^+$ transition. The fits (red solid lines) include peaks (blue dashed line) corresponding to the 0~keV ($0_{1}^+$), 1229 ($0_{2}^+$), 1465 ($0_{3}^+$) and 2084~keV ($1^+,2^+$) $^{96}$Sr states, as well as a continuous background for the total spectrum. Note that for the $0^+_3$ state the $\gamma$-ray detection efficiency is lower due to the 6.7(10) ns half-life and the 38\% branching ratio to the $0_2^+$ state~\cite{ENSDF}. }
\label{ExcSpec}
\end{figure}

\begin{figure}[h!]
\includegraphics[width=\columnwidth]{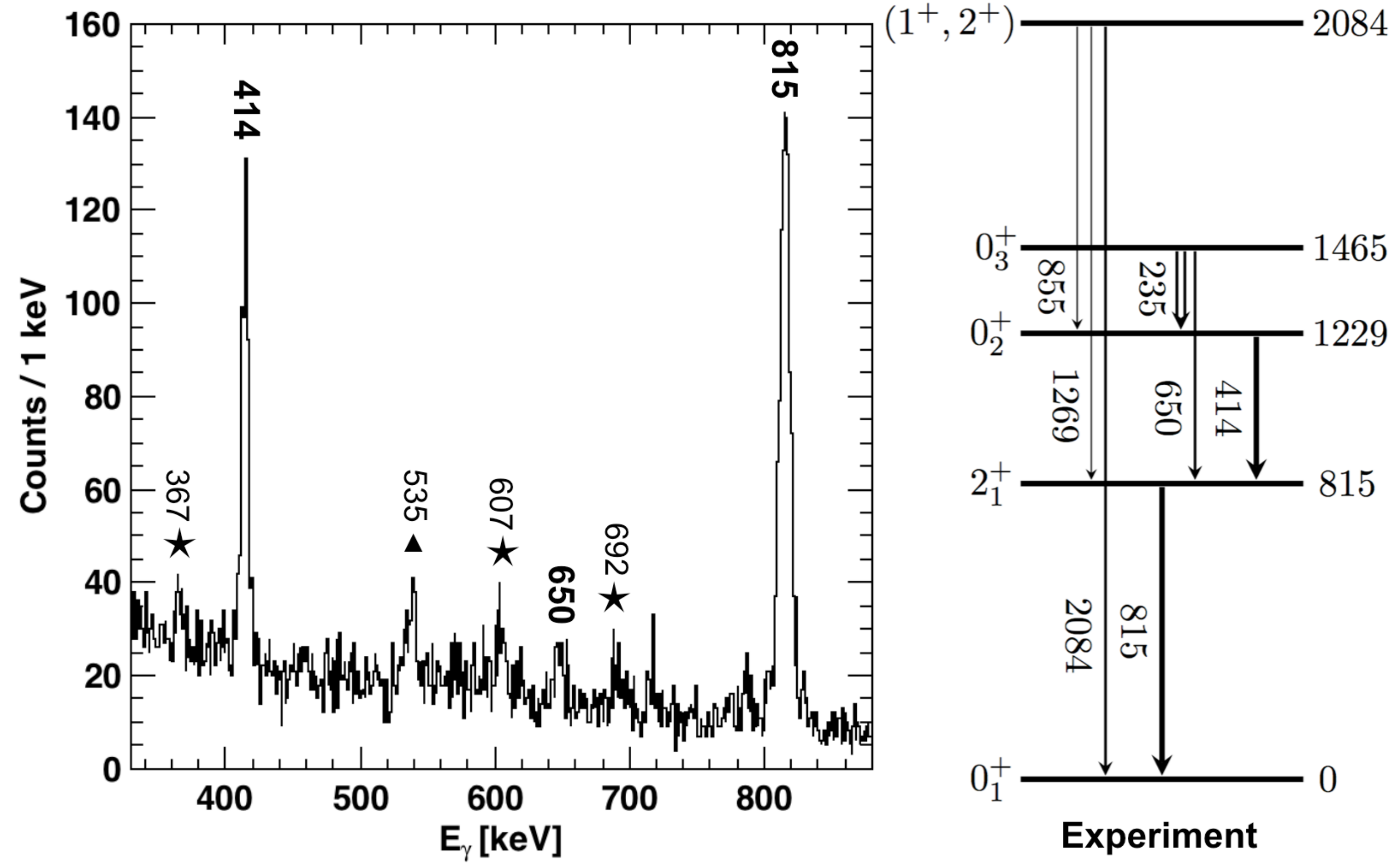}%
\caption{$\gamma$-ray energy spectrum coincident with $^{96}$Sr states in the range 800 $< E_\text{x} <$ 1900~keV. Only TIGRESS detectors positioned at $\theta_\text{lab}>120^\circ$ were used. Transitions marked with a star are from the decay of the 1507, 1995, 2084 and 2113~keV states which were partially included in the excitation energy gate. These states will be discussed in detail in a forthcoming publication. The transition marked with a triangle could not be identified. A partial experimental level scheme is shown on the right, including only states and transitions relevant for this work as well as the 235~keV $E0$ transition.}
\label{Gammas}
\end{figure}

The excitation energy of states in $^{96}$Sr populated through the \dpsr\ reaction was determined by measuring the proton energy and scattering angle. 
Figure~\ref{ExcSpec} shows two reconstructed $^{96}$Sr excitation energy ($E_x$) spectra produced using only data from backwards laboratory angles ($\theta_\text{lab}>90^\circ$) as the excitation energy resolution is improved in this angular range. 
The direct population of the $0_1^+$ $^{96}$Sr ground state is clearly visible while the large $\beta$-decay background (from accidentally stopped \nuc{95}{Sr}) and density of states made it impossible to resolve higher lying states. Excited states were thus identified using the de-excitation $\gamma$-ray in addition to an $E_\text{x}$ gate.
 A $\gamma$-ray gate on the 414~keV $0_{2}^+ \rightarrow 2_1^+$ transition was used to produce the $\gamma$-ray efficiency-corrected excitation energy spectrum (open points). In this spectrum, the 1229~keV ($0_{2}^+$), 1465~keV ($0_{3}^+$) and 2084~keV ($1^+,2^+$) $^{96}$Sr states could be resolved using a fit  (blue dashed lines) that utilized a fixed resolution (FWHM = 425~keV), determined from the ground state, and the known excitation energies of the states.
Figure~\ref{Gammas} shows the Doppler-reconstructed $\gamma$-ray energy spectrum for TIGRESS detectors positioned at backward angles ($\theta_\text{lab}>120^\circ$) in coincidence with protons. An excitation energy gate of 800$<E_\text{x}<$1900~keV was utilized 
to include protons associated with the direct population of the 1229~keV ($0_{2}^+$) and 1465~keV ($0_{3}^+$) states.

For the Doppler reconstruction, the $\gamma$ rays were assumed to be emitted at the center of the target. This is not correct for the 650~keV transition from the long-lived (6.7 ns~\cite{ENSDF}) 1465~keV $0_{3}^+$ state, which decays outside the target. By using only the most backward TIGRESS detectors a reasonably narrow Doppler-corrected peak could be achieved for this transition, since here the difference between the real emission angle of the $\gamma$ ray and the angle assumed for the Doppler reconstruction is sufficiently small even for a relatively long-lived state. This allowed for a clear identification of this transition, despite the reduced $\gamma$-ray detection efficiency.
The cross-section for the population of the 1465~keV $0_{3}^+$ state was determined using a relative $\gamma$-ray intensity analysis between the 650~keV $0_{3}^+\rightarrow 2_{1}^+$ line and the 414~keV $0_{2}^+\rightarrow 2_{1}^+$ line (Figure~\ref{Gammas}). The experimentally observed yields for these transitions were compared to a detailed \textit{Geant4}~\cite{GEANT} simulation of the decay of the 1229 and 1465~keV states, taking into account the TIGRESS geometry, attenuation of $\gamma$-rays in the SHARC chamber and beam-line material, the kinematics of the recoiling $^{96}$Sr nucleus ($\beta\approx0.1$) and the known half-lives and decay branching ratios of the two states. From the ratio of the number of measured $\gamma$ rays it was deduced that the relative population strength of the 1465~keV state compared to the 1229~keV state is 1.50(52).

\begin{figure}[h!]
\includegraphics[width=\columnwidth]{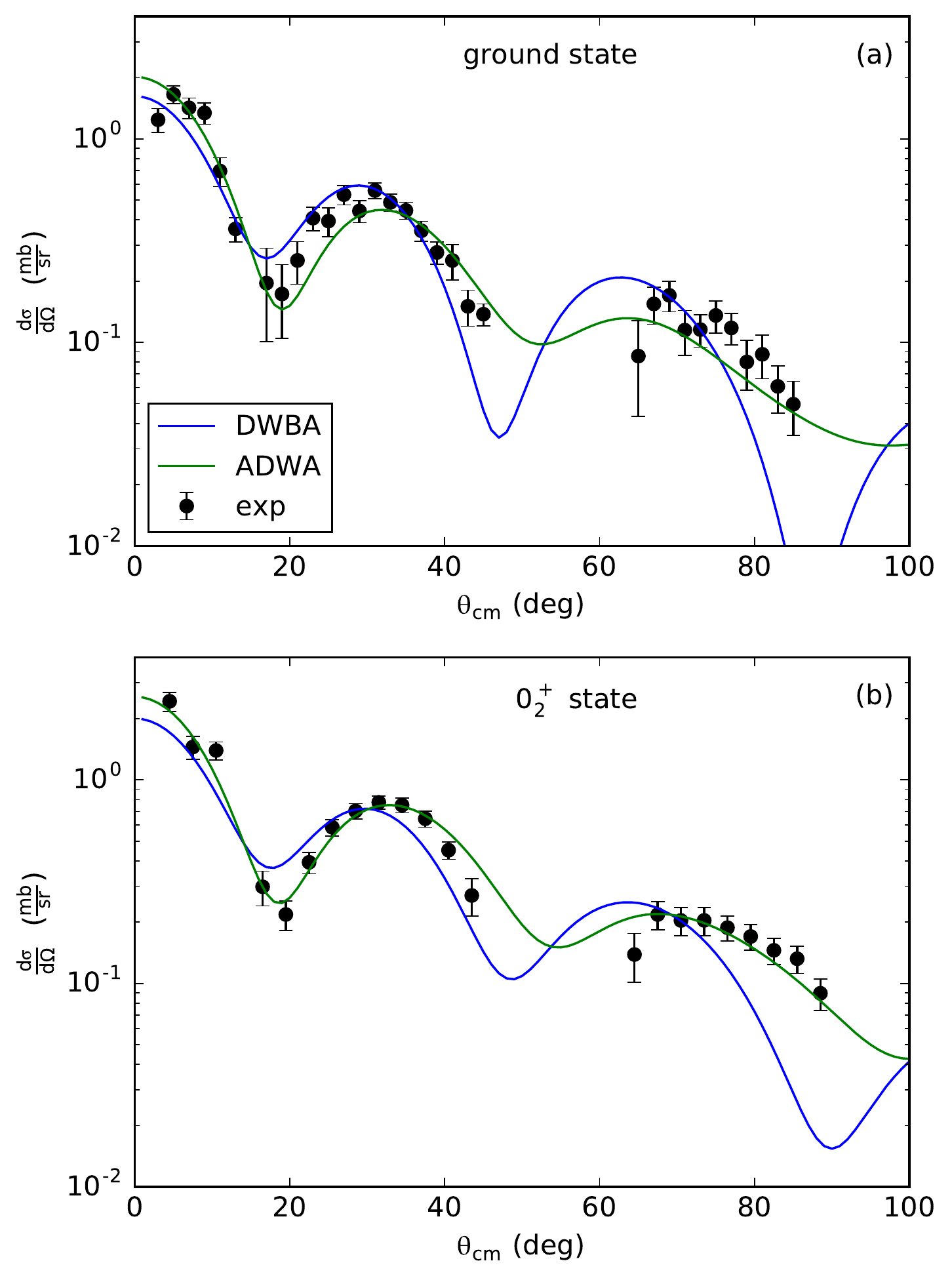}%
\caption{Fit of DWBA (blue) and ADWA (green) calculations to experimental data for the $^{96}$Sr $0_{1}^+$ ground state (a) and 1229~keV $0_{2}^+$ state (b). For the determination of the spectroscopic factors only the forward center-of-mass angles $\theta_\text{cm}<45^\circ$ were considered.}
\label{AngDist}
\end{figure}

Figure~\ref{AngDist} shows the experimental angular distributions for the $0_{1,2}^+$ $^{96}$Sr states compared to distorted wave Born approximation (DWBA) and adiabatic distorted wave approximation (ADWA) calculations that were carried out using \textit{FRESCO}~\cite{FRESCO}.
For the DWBA calculations, the optical model (OM) parameters were optimized using the elastic scattering angular distributions for d($^{95}$Sr,d) and p($^{95}$Sr,p), the data for which were acquired simultaneously with d($^{95}$Sr,p). For d($^{95}$Sr,d) the OM parameters of Lohr and Haeberli~\cite{LHOpticalModel} were used, with slight adjustments to better reproduce the elastic scattering data. The p($^{95}$Sr,p) data was dominated by pure Rutherford scattering and was, within uncertainties, not sensitive to different OM parameters. These calculations also provided the normalization of the cross section. The analysis procedure of Wilson \textit{et. al.}~\cite{Wilson2016} was followed, and further details will be provided in a forthcoming publication~\cite{Cruz2018}. For the ADWA calculation global nucleon-nucleus OM parameters from~\cite{koning03} were used.
The calculations (Figure~\ref{AngDist}) model the \dpsr\ reaction as a single-step process where the transferred neutron populates the $\nu2s_{1/2}$ orbital via pure $\ell=0$ angular momentum transfer. The ADWA calculations better reproduce the data at large scattering angles where deuteron breakup is expected to affect the transfer cross section. 
The spectroscopic factors $C^2S$ are determined as the ratio of experimental to reaction model cross section, the fits were restricted to the forward angles $\theta_\text{cm}<45^\circ$. For the transfer to the $0_{1,2}^+$ $^{96}$Sr states the spectroscopic factors amount to $C^2S=0.19(3)$ and 0.22(3), respectively, for the DWBA and 0.15(3) and 0.19(3) for the ADWA calculations. The spectroscopic factor uncertainties include both statistical and systematic contributions related to the normalization of the data. Additional uncertainties for the absolute value of the spectroscopic factors arise from the choice of the reaction model and the optical model potential parameters especially for the DWBA. Based on the comparison of the two reaction models and various sets of optical model parameters these amount to 20~\%. For the relative spectroscopic factors of the \zps\ used to extract their mixing strength these uncertainties cancel. The inclusion of multi-step processes through coupled channels calculations leads to a slightly better description of the differential cross section, at the expense of additional unconstrained parameters. Complete and detailed coupled-channel calculations are beyond the scope of the present work. To estimate the contribution, the inelastic excitation of \nuc{95}{Sr} was taken into account through an effective deformation length. A value of $\delta_n=1.1$~fm reproduces the measured $(d,d^\prime)$ cross section and is consistent with the spherical nature of \nuc{95}{Sr} determined from the charge radius~\cite{Buchinger1990} and the neighboring \nuc{94,96}{Sr}~\cite{Clement2016a,Regis2017}. Changes to the spectroscopic factors are less than 10~\% and do not alter the conclusion presented in the following.

It was not possible to extract an angular distribution for the 1465~keV $0_3^+$ state due to low statistics. However, from the analysis of the relative population strength discussed above, the spectroscopic factor for the 1465~keV state could be deduced to be 0.33(12) or 0.29(13), for DWBA and ADWA calculations, respectively. The relative spectroscopic factors used in the discussion below are not affected by the choice of the reaction model and the systematic uncertainties arising from the normalization of the cross section.

The relative strengths of the $0_{2,3}^+$ states in $^{96}$Sr were interpreted using a two-level shape mixing model. The $0_1^+$ ground state of $^{96}$Sr was excluded from the mixing model analysis as the mixing of this state with the excited $0_{2,3}^+$ states is expected to be negligible. This is evidenced by the fact that no $E0$ transitions between the excited $0_{2,3}^+$ states and the $0_1^+$ ground state of $^{96}$Sr were reported in the work of Jung~\cite{JungThesis}. This assumption is also supported by the recent Coulomb excitation data and beyond mean field calculations reported by Cl\'{e}ment \textit{et al.}~\cite{Clement2016,Clement2016a}. Using the two-level mixing model, the monopole transition strength between the $0_{2,3}^+$ $^{96}$Sr states is related to their mixing strength $a^2$ and intrinsic quadrupole deformations $\beta$ by, 
\begin{equation} \label{eqn:monopole}
\rho^2(E0) = \bigg( \frac{3}{4\pi} \bigg)^2 Z^2  a^2(1-a^2) \big[\Delta(\beta^2)\big]^2
\end{equation}
where $Z$ is the atomic number~\cite{Wood1999}. 
In the case where the unmixed states are a spherical configuration $0_\text{sph}^+$ ($\beta_\text{sph}=0$) and a strongly deformed configuration $0_\text{def}^+$ (with $\beta_\text{def}$), the difference in (squared) deformation between the configurations is $\Delta(\beta^2)=\beta_\text{def}^2$. 
The wave functions of the $0_{2,3}^+$ states in \nuc{96}{Sr} would therefore be $\ket{0_2^+} = a \ket{0^+_{\text{sph}} } + \sqrt{1-a^2}  \ket{0^+_{\text{def}} }$ and $\ket{0_3^+} = \sqrt{1-a^2} \ket{0^+_{\text{sph}} } - a \ket{0^+_{\text{def}} }$, respectively, with the mixing amplitude $a$. 
 Given that the ground state of $^{95}$Sr has a nearly spherical shape~\cite{Buchinger1990}, a substantial re-arrangement of the valence nucleons would be required in order to directly populate a strongly deformed configuration in $^{96}$Sr through a single-step transfer reaction such as (d,p).
 It is therefore assumed that there was negligible direct population of the deformed configuration, $0_\text{def}^+$, in this experiment and so the strength of the $0_{2,3}^+$ states in $^{96}$Sr is a direct measure of the $0_\text{sph}^+$ content of the excited 1229 and 1465~keV state wave functions. The ratio of the spectroscopic factors of the $0_2^+$ and $0_3^+$ states in $^{96}$Sr is therefore equal to $\frac{1-a^2}{a^2}$, giving $a^2=0.40(14)$. This result is independent of the reaction model choice as the ratio is determined by the ratio of cross sections with a dynamical correction accounting for the difference in excitation energy.
 By combining the known value of $\rho^2(E0)$ with our experimental constraint on the mixing strength $a^2$, equation~\ref{eqn:monopole} was used to determine the absolute value of $\beta_\text{def}=0.31(3)$. 
In this strong mixing scenario the interaction strength between the $0_{2,3}^+$ states in $^{96}$Sr is 113~keV and the energies of the unmixed $0^+_{\text{def}}$ and $0^+_{\text{sph}}$ states are 1314 and 1380~keV, respectively.
It is interesting to compare these results to the shape coexistence in neighboring $^{98}$Sr. Here, the $0_2^+$ state is situated only 215~keV above the ground state and a two-level mixing model resulted in only a weak mixing between the coexisting states~\cite{Clement2016a,Park2016}. As a result, the $^{98}$Sr $0_1^+$ ground state is strongly deformed whereas the excited $0_2^+$ is nearly spherical. The weak mixing also implies a surprisingly small interaction strength between the unmixed configurations of only $\approx 10$~keV.

The strongly populated 2084~keV state (Figures~\ref{ExcSpec} and~\ref{Gammas}), has a 51~\% branching ratio via the 855~keV transition to the 1229~keV $0_2^+$ state relative to the 2084~keV ground state transition. Using the $0_{2,3}^+$ mixing strength of 0.40(14) one can calculate the expected branching ratio for the 2084~keV to 1465~keV $0^+$ state transition if we assume that the transition rate to the unmixed $0^+_{\text{def}}$ is negligible. Based on this, the branching ratio for this 619~keV transition is expected to be $57^{+51}_{-25}$ ($30^{+27}_{-13}$) \% relative to the 2084~keV ground state transition within $1\sigma$ uncertainties, assuming that the 2084~keV state has a spin and parity of $1^+$ ($2^+$). 
No measurement of this transition has been reported~\cite{ENSDF} and it was not observed in the present experiment. Thus, the observation of the 855~keV transition and the non-observation of a hypothetical 619~keV indicates that the 1229~keV state contains a larger component of the $0^+_\text{sph}$ configuration than the 1465~keV state ($a^2 > 0.5$). At the same time the $E0$ and $E2$ branching ratios and the very different half-lives of the isomeric 1465~keV state and the 1229~keV state favor $a^2 > 0.5$ as well.

Shell model calculations were carried out using \textit{NushellX}~\cite{NushellX}, employing the \textit{glek} interaction~\cite{glek}. The model space comprises of the proton $fpg_{9/2}$ and neutron $gds$ orbitals. The two-body matrix elements are obtained from G-matrix calculations with some modifications to better describe the Y and Zr nuclei~\cite{glek}. For the calculations presented here, the single-particle energies were adjusted to reproduce low energy states for odd mass nuclei in the vicinity of $Z=40$ and $N=58$. 
States in $^{95,96}$Sr and spectroscopic factors for \dpsr\ were calculated using several different valence spaces to investigate the influence of the various proton degrees of freedom. For the neutrons an inert $N=50$ core was assumed and the valence space included the $\nu [2s_{1/2}]$, $[1d_{3/2}]$, $[1d_{5/2}]$, and $[0g_{7/2}]$ orbitals. Calculations were carried out separately using three different proton valence spaces.  In valence space \textcircled{a}, the protons are required to be inert in a $\pi [1p_{3/2}]^4$ configuration. In valence space \textcircled{b}, proton excitations into the nearby $\pi [1p_{1/2}]$ orbital are allowed. Finally, valence space \textcircled{c} further expands the proton valence space to include the $\pi [0g_{9/2}]$ orbital by allowing for two additional proton excitations across the $Z=40$ sub-shell gap. The occupancy of the $\pi [0g_{9/2}]$ orbital was restricted to two protons due to computational limitations. 
Table~\ref{ShellModelSF} compares the experimental spectroscopic factors for the $0^+$ states with those predicted by the shell model calculations for the three different configuration spaces. 
\begin{table*}[h]                 
\begin{center}
 \caption{Comparison of experimental (using the DWBA calculations) to calculated spectroscopic factors ($C^2S$) for $0^+$ states in $^{96}$Sr populated via the \dpsr\ reaction. The experimental values for the two excited $0^+$ states result in the unmixed spherical state, while for the shell model calculations the two $0^+$ states with the highest $C^2S$ are listed (for details see text).}
  \label{ShellModelSF}
  \begin{tabular}{cc|cc|cc|cc|cc} 
    \hline
    \hline
     \multicolumn{2}{c}{Exp.} \vline & \multicolumn{2}{c}{Unmixed} \vline & \multicolumn{2}{c}{\textit{glek} \textcircled{a}} \vline & \multicolumn{2}{c}{\textit{glek} \textcircled{b}} \vline & \multicolumn{2}{c}{\textit{glek} \textcircled{c}} \\
    $E_\text{x}$ (keV) & $C^2S$ & $E_\text{x}$ (keV) & $C^2S$  &  $E_\text{x}$ (keV) & $C^2S$ & $E_\text{x}$ (keV) &  $C^2S$ & $E_\text{x}$ (keV) & $C^2S$  \\
    \hline
    0    & 0.19(3)  & 0    & 0.19(3)  &    0    & 1.742 &   0     &1.575 &  0     & 1.455 \\ 
    1229 & 0.22(3)  & 1314 & 0        & - & - & - & - & - & - \\
    1465 & 0.33(13) & 1380 & 0.55(13) & 2271 & 0.056 & 1691 & 0.098 & 444 & 0.105 \\
    \hline
    \hline    
  \end{tabular}
 \end{center}
\end{table*}
The experimental value for the unmixed 1380~keV $0^+$ state shown in Table~\ref{ShellModelSF} corresponds to the total excited $0_\text{sph}^+$ spectroscopic factor, the sum of the experimental spectroscopic factors for the $0_{2,3}^+$ states. The shell model calculations predict a much larger neutron $\nu [2s_{1/2}]^2$ component for the ground state than the excited $0^+$ state while experimentally the opposite is observed. Including the $\pi [0g_{9/2}]$ proton configuration in the shell model calculations only slightly increases the population of the excited $0^+$ state, while dramatically lowering its energy (see Table~\ref{ShellModelSF}). Experimentally, about half of the overall predicted $\nu [2s_{1/2}]^2$ strength is observed in the low-lying $0^+$ states.
The present shell model interactions and single-particle energies also describe well the experimental spectroscopic factors for the low-lying states in $^{95}$Sr populated via the d($^{94}$Sr,p) reaction, carried out as part of the same experimental campaign. For the $1/2^+$ ground state of \nuc{95}{Sr} a $C^2S=0.45$ (for valence space \textcircled{b} which described best the excitation energies) is predicted while the measurement using the same analysis as presented here results in $C^2S=0.41(9)$. Further results on the d($^{94,95}$Sr,p) reactions will be reported in a forthcoming publication~\cite{Cruz2018}.

 These results show, that while the \nuc{94,95}{Sr} nuclei as well as the strongly deformed \nuc{98}{Sr} nucleus can be described rather well using shell model~\cite{Regis2017,Cruz2018} and beyond mean field calculations~\cite{Clement2016,Clement2016a}, \nuc{96}{Sr} at $N=58$, just before the shape transition, has a much more complicated structure. The present transfer reaction study enables to selectively populate the spherical component of the $0^+$ states in \nuc{96}{Sr}. The spherical component is found mainly in the strongly mixed excited $0^+$ states. The mixing ratio was determined for the first time in the present study. The ground state on the other hand is only weakly populated in the \dpsr\ reaction suggesting a triple shape coexistence in \nuc{96}{Sr} with a (weakly) oblate or triaxial ground state.

In summary, we have measured the population of low-lying states in $^{96}$Sr via the \dpsr\ reaction at 5.5 $A$MeV. 
The results show a surprisingly strong population of an excited spherical configuration in $^{96}$Sr, which itself is strongly mixed with a deformed ($\beta$  = 0.31(3)) configuration, giving rise to two $0^+$  states at 1229 and 1465~keV. Owing to the almost pure $\nu [s_{1/2}]$ ground state configuration of the $1/2^+$ ground state in $^{95}$Sr, the (d,p) transfer is mostly sensitive to the $\nu [s_{1/2}]^2$ configuration in the final $0^+$ state. Contrary to the experimental data, shell model calculations using a constrained model space predict a predominant $\nu [s_{1/2}]^2$ configuration for the \nuc{96}{Sr} ground state. This suggests the occurrence of three distinct shapes in \nuc{96}{Sr}.
Clearly, more extensive theoretical studies are required to gain better insights into the single-particle wave functions in the Sr isotopes in this context.  Extensions of the work carried out for the Zr isotopes with large scale shell model calculations~\cite{Sieja2009} and Monte Carlo Shell Model calculations~\cite{Togashi2016}, as well as further developments of the beyond mean field calculations~\cite{RodGuz2010,RodGuz2010a,Clement2016,Clement2016a}, will be of interest. 

The efforts of the TRIUMF operations team in supplying the \nuc{95}{Sr} beam are highly appreciated. We acknowledge support from the Science and Technologies Facility Council (UK, grants EP/D060575/1 and ST/L005727/1), the National Science Foundation (US, grant PHY-1306297), the Natural Sciences and Engineering Research Council of Canada, the Canada Foundation for Innovation and the British Columbia Knowledge and Development Fund. TRIUMF receives funding via a contribution through the National Research Council Canada.

\bibliographystyle{elsarticle-num-names}
\bibliography{references}

\end{document}